\documentstyle[12pt]{article}
\begin{document}
\begin{titlepage}
\begin{center}
\null
\vskip-1truecm  
{\footnotesize Available at: 
{\tt http://www.ictp.trieste.it/\~{}pub$_-$off}}\hfill IC/98/36
\vskip1truecm
United Nations Educational Scientific and Cultural Organization\\
and\\
International Atomic Energy Agency\\
\medskip
THE ABDUS SALAM INTERNATIONAL CENTRE FOR THEORETICAL PHYSICS\\
\vskip2truecm
{\bf LOCALIZATION AND ABSORPTION OF LIGHT\\
IN 2D COMPOSITE METAL-DIELECTRIC FILMS \\
AT THE PERCOLATION THRESHOLD}\\
\vspace{1.5cm}
L. Zekri\\ 
{\em U.S.T.O., Departement de Physique, L.E.P.M.,\\  B.P.1505 El 
M'Naouar, Oran, Algeria\\
and\\
The Abdus Salam International Centre for Theoretical Physics,  Trieste, Italy,}\\
\bigskip
R. Bouamrane\\ 
{\em U.S.T.O., Departement de Physique, L.E.P.M.,\\  B.P.1505 El 
M'Naouar, Oran, Algeria,}\\
\bigskip
N. Zekri\footnote{\normalsize Regular Associate of the ICTP.}\\
{\em U.S.T.O., Departement de Physique, L.E.P.M.,\\  B.P.1505 El 
M'Naouar, Oran, Algeria\\
and\\
The Abdus Salam International Centre for Theoretical Physics,  Trieste, Italy}\\
\bigskip
{\em and}\\
\bigskip
F. Brouers\\
{\em Universite de Liege, Institut de Physique, Etude 
Physique des Materiaux, \\
Sart Tilman 4000, Liege Belgium. } \\
\vspace{2cm}
MIRAMARE -- TRIESTE\\
\medskip
April 1998
\end{center}
\end{titlepage}

\centerline{\bf Abstract}
\baselineskip=20pt

We study in this paper the localization of light and the dielectric properties of thin 
metal-dielectric composites at the percolation threshold and around a resonant
frequency where the conductivities of the two components are of the same order.
In particular, the effect of the loss in metallic components are examined. To this end,
such systems are modelized as random $L-C$ networks, and the local field distribution as
well as the effective conductivity are determined by using two different methods for
comparison: an exact resolution of Kirchoff equations, and a real space renormalization
group method. The latter method is found to give the general behavior of the effective
conductivity but fails to determine the local field distribution. It is also found  that the
localization still persists for vanishing losses. This result seems to be in agreement with
the anomalous absorption observed experimentally for such systems. 

\newpage

\noindent{\bf Introduction }

\hspace{0.33in}
Thin metal-dielectric films have been shown experimentally to exhibit an anomalous
high absorption in the visible, near infrared and microwave regimes and near the 
percolation
threshold[1-4]. This effect was interpreted as cluster plasmon absorption 
\cite{Sary1,Brou1}.
Naturally, these mixtures should show an absorption since the dielectric constant of the
metallic component is complex and the effective dielectric constant of the whole
system should then also be complex particularly at the percolation threshold where 
\cite{Dykh}
\begin{equation}
\epsilon_{eff} =\sqrt{\epsilon_{m} \epsilon_{d}} \label{effective} \end{equation}
The indices $m, d$ and $eff$ stand respectively for the metal, dielectric and effective 
medium. From this equation, both real and imaginary parts of the dielectric
constant of the metallic component contribute to the optical absorption of the system.
Furthermore, a vanishing metallic absorption naturally should lead to a vanishing
absorption of the whole system since a superconductor-dielectric mixture should be non
dissipative. Therefore, this description of the effective properties of such medium
may not be sufficient to explain the behavior experimentally observed for such systems
[1-4].

\hspace{0.33in} Indeed, since the light has a wave behavior, the backscattering and the
interference effects can strongly affect its propagation through a disordered system
\cite{John,Lee,Gena} and its localization is enhanced by disorder \cite{Lee} as
well as by absorption \cite{Gupt}. Therefore, the localization properties of the
electromagnetic field in such systems where both absorption and disorder are present, can
be a good tool for explaining the anomalous behavior observed in such thin composite
films, particularly at the classical percolation threshold which is a transition point
of the effective $dc$ conductivity of the system from non-conductor to conductor due to
the appearance of a continuous path of the conducting region through the sample. The
classical percolation threshold in 2D disordered bonds corresponds to a concentration of
the metallic bonds $p_{c}=0.5$ \cite{Berg} (note that some composites carry current
even below the classical percolation threshold due to the fact that tunnelling through
disconnected (dispersed) metallic regions can give some virtually connected percolating
clusters \cite{Frit}).

\hspace{0.33in} On the other hand, it has recently been  found numerically in such
films, giant local field fluctuations \cite{Brou2} at the percolation threshold and for 
frequencies close to a resonant one $ \omega_{res} $ where the conductivities
of the two components are of the same order ($\left| \sigma_{m} \right| = \left| \sigma_{d} 
\right|$). High local field fluctuations have also been found both in fractal 2D films 
\cite{Stoc}, 3D rough
surfaces \cite{Mark} and non-linear Raman scattering \cite{Shal}. In both systems the
electromagnetic modes were found to be localized, apparently, due to such fluctuations.
Furthermore, Brouers et al. \cite{Brou2} showed for the metal-dielectric films that the local 
field distribution is asymptotically log-normal. However, from the
electromagnetic field theory investigated recently by Sarychev et al. \cite{Sary1},
we can easily deduce that the high strengths of the current (or equivalently the high
local field intensities in this case) behave as the inverse of the local transmission
of light. We deduce, then, that the local transmission has also a log-normal distribution in 
such films. Therefore, if we use the analogy between the electric
field in Helmoltz equation and the electronic wave-function in Schr\"odinger equation
\cite{John}, the local transmission is equivalent to the electronic conductance at
zero temperature \cite{Land} where a log-normal distribution is a signature of localization 
\cite{Chas,Seno}.

\hspace{0.33in} This is the aim of the present paper where we study the localization
and absorption properties of such films modelized by a square $RL-C$ network. The
local field and the effective conductivity are calculated by using two different
methods for comparison with the results of \cite{Brou2}: an exact resolution of the
Kirchoff equations for such network which we call Exact Method (EM) from now on, and a
Real Space Renormalization Group (RSRG) self-similar scheme \cite{Sary2}. The degree
of localization is measured by means the inverse participation ratio (IPR) \cite{Bell}
applied to the electric field while the absorption is deduced from the real part of
the effective conductivity. We compare in a first step the field distribution obtained by the 
two methods and then examine the effect of the loss in the metallic
component on the localization as well as the absorption at the percolation threshold.

\noindent{\bf Method of the calculations } 

\hspace{0.33in}
The conductivity of metallic and dielectric grains is related to their dielectric
constant by \cite{Sary1}
\begin{equation}
\sigma_{m,d}= \frac{-i d \omega \epsilon_{m,d}}{4 \pi} \label{sigmaepsi} 
\end{equation}
where $d$ is the film thickness, $\omega$ the field frequency and $\epsilon_{m,d}$
the dielectric constants respectively of the metal and dielectric components assumed to be 
homogeneous spheres. Here the film thickness and the size of the components must be 
smaller than the light wavelength in order to neglect the magnetic field variation. The 
dielectric constant of the insulator is real while
the metallic one is complex and, from Eq.(2), its imaginary part (absorption) is
related to the real part of the conductivity. 

\hspace{0.33in} When the light frequency is large compared to the relaxation frequency, 
composite metal dielectric films can be modelled by 2D resistor networks
\cite{Sary1,Brou1,Zeng}. The effective properties of these networks have already been
extensively studied during the last two decades [5,6,23-25]. In the $RL-C$ picture,
and if the frequency is smaller than the plasmon frequency $\omega_{p}$, capacitors
$C$ stand for the dielectric grains with a conductivity $\sigma_{C} = -iC \omega $
and a concentration $1-p$ while inductances $L$ represent the metallic grains with a
conductivity $\sigma_{L} =(-iL \omega + R)^{-1}$ ($R$ being the loss), and a 
concentration $p$ deposited or evaporated over the substrate. Therefore we can take in
this case, without loss of generality, $L = C = \omega_{res} =1 $ (the resonant
frequency $ \omega_{res} $ corresponds to the frequency where metallic and dielectric
conductivities have the same magnitude for small losses, i.e., $ C \omega_{res} = 1/ L 
\omega_{res}$ ). $L$ and $C$ are constants near the resonant
frequency, this assumption can be generalized to any frequency $ \omega $ (smaller than 
the
plasmon frequency) normalized to $\omega_{res}$. The metallic conductivity  for 
small
losses becomes
\begin{equation}
\sigma_{L}= \frac{1}{-i\omega + R} = ( i + \frac{R}{\omega})/ \omega \end{equation}
while the dielectric conductivity is
\begin{equation}
\sigma_{C}=-i \omega
\end{equation}

\hspace{0.33in} The first method (EM), used for the calculation of the local field
distribution and the effective conductivity, consists in solving exactly Kirchoff
equations for the corresponding 2D square resistor network. This implies the use of
$M^{2}$ x $M^{2} $ matrices (where $M$ is the size of the square lattice), which are
impossible to handle numerically for large samples (for memory and computational time
consuming reasons). However, we take advantage of the sparce configuration of such
matrices and their organization in blocks in their diagonal region. The diagonalization and
inversion of such matrices, then, is obtained simply from the diagonalization and inversion 
of
the constituent blocks. This reduces considerably the memory and the computational time
consuming. We note that this method provides exactly the same results on the effective
conductivity as the Frank and Lobb method \cite{Frank} but calculates also the local
field distribution through the lattice which cannot be done by the other method. However, 
the
time consuming remains large with this method (in particular when averaging over a large 
number
of configurations), and the maximum size we reach by this method is $256$ x $256$ 
(which is
sufficient for our statistical treatment). 

\hspace{0.33in}
This is one of the reasons for using also the RSRG method which is much less 
computational time
consuming. This method, extensively described in previous works (see 
\cite{Brou2,Sary2,Brou3}),
consists in a transformation of the 2D square lattice into Weatstone bridges in $x$ and $y$
directions (see Fig.1). Each bridge is transformed into an equivalent admittance, and after a
number of steps the lattice is reduced into two equivalent admittances following these 
directions.
It is then easy to calculate by this transformation the effective conductivity while the
local field distribution can be obtained by the inverse procedure starting from the effective
admittances already calculated. Although this method is an approximation, it has been 
shown to
give values of the effective conductivity near the percolation threshold very close to the
exact ones for 2D composites \cite{Dykh,Brou3} and critical exponents not far away from 
the
known values of the percolation theory \cite{Berg}. Furthermore, this method uses only 
few
matrices of $M$x$M$ for sample sizes $M$x$M$ which reduces considerably the 
computational memory
in comparison with the other methods. We can, then, easily reach sizes of $1024$x$1024$ 
with the
same computer configuration as for the previous method. However, the first method
(EM) is also needed in the present work since the validity of RSRG in calculating the local
field has not been checked before.

\hspace{0.33in}
As discussed in the previous section, it seems that the localization properties of the
optical waves in thin metal-dielectric composites is an interesting way to explain the
anomalous absorption observed near the percolation threshold. This classical threshold
corresponds to the appearance of an infinite metallic channel which, in 2D disordered
systems, is reached for an equal probability of the two components \cite{Berg}. One of
the useful quantities to study the localization in electronic systems is the inverse
participation ratio (IPR) \cite{Bell}. By analogy with the quantum counter part, the
local electric field in Helmoltz equation plays the role of the electronic wave function in
the Shr\"odinger equation \cite{John} and the IPR becomes 
\begin{equation}
IPR = \frac{\Sigma_{i} \left| E_{i} \right|^{4}}{\left( \Sigma_{i} \left| E_{i} \right|^{2}
\right)^{2}}
\end{equation}
where $E_{i}$ denotes the local electric field at site $i$. The IPR has been defined for the
electronic waves in order to measure the spatial extend of the dominant eigenstates and to
characterize the electronic states in disordered materials \cite{Bell}. Therefore, for the
electromagnetic eigenmodes this quantity will behave as 
\begin{eqnarray}
IPR = O(M^{-d}) \,\,\,\;\,\,\, for \,\, extended \,\, eigenmodes, \\ IPR = O(M^{0}) 
\,\,\,\;\,\,\, for \,\, strongly \,\, localized \,\, eigenmodes.
\end{eqnarray}
Here $d$ denotes the Euclidian dimension of the system ($d=2$ in this case) and $M$
the size of the system. Thus in the case of purely extended eigenmodes, the field has a
significant strength over the whole surface of the film and the denominator will be
$M^{4} \left| E \right|^{4}$ while the numerator behaves as $M^{2} \left| E \right| ^{4}$
(by assuming the field constant) leading to a decay of the IPR as $M^{-2}$. In the case of
strongly localized eigenmodes, the more significant field strengths are located in a limited
area of average size $M_{c}^{2}$ where $M_{c}$ is the localization length. It is then 
obvious
that the IPR remains constant outside this area. Therefore we can estimate the degree of
localization of the light from the power-law decay exponent of the IPR which varies from 
$0$
(for strongly localized eigenmodes) to $-2$ (for purely extended eigenmodes). This 
exponent is
determined by the slope of the variation of the IPR as a function the system size in log-log
scale. It also measures the correlation fractal dimension ($-D_{2}$) of the local field
\cite{Brou2,Mato}.  From now on we will call this exponent the slope of the IPR.

\noindent{\bf Results and Discussion }

\hspace{0.33in}
In this section we consider a film of size $256$x$256$. As discussed above, the RSRG 
method is a
good approximation for the calculation of the effective conductivity and the critical 
exponents.
However, this agreement cannot be generalized to any other quantity. In particular, the 
value of
the effective conductivity at the percolation threshold is due only to the appearance of an
infinite metallic cluster channel in the sample no matter how the bonds are distributed
over the sample \cite{Efro,Berg}. However, the distribution of bonds can affect
sensitively the local field distribution which is the main quantity to measure the
localization properties of light in this system. Thus, some particular arrangements of the
bonds yield a local field intensity ($\left| E \right| ^{2}$) of the order of $R^{-2}$ by the
RSRG method. Therefore, in the limit of vanishing losses the field intensity will diverge
which is unphysical. This divergence can limit the validity of the RSRG method for vanishing 
losses.

\hspace{0.33in}
In Figs.2 we compare the distributions of the local field intensity obtained by the RSRG 
method
to those of the EM for sample sizes $256$x$256$, for two different losses and two
different frequencies in order to check the validity of RSRG in calculating the local
field. The RSRG distributions seem to be wider, with very large field strengths, than for the
EM distributions (supporting the previous discussion on the divergence of the field) which
seem to be perfectly log-normal for any loss and frequency. Furthermore, other peaks of
small field strengths appear in the distribution obtained by RSRG particularly for small
losses and at the resonant frequency $\omega = 1$, while for higher losses and different
frequencies, these peaks move to larger field strengths and overlap with the main one
contributing to the broadening of the distribution. This behavior affects strongly the
IPR calculations since the broadening of the distribution means an increase of the 
localization \cite{Chas,Seno}. However, the additional peaks appearing
for small losses in the RSRG method should contribute only slightly to the IPR since they are 
many
orders smaller than the main peak. Therefore, although the ditributions of field for RSRG
are clearly different from those of the EM, it seems that the results for $\omega = 1$
provide the best fit for the localization properties of these films. Therefore, we will
restrict ourselves to the resonant frequency ($\omega = 1$) from now on. 

\hspace{0.33in}
In Figs.3, we show, by using both RSRG and EM methods, the real part of the 
conductivity averaged over $50$ samples of size $256$x$256$ (Fig.3a) as well as the 
slope
of the IPR (Fig.3b) as a function of the loss parameter $R$ for $\omega =1$. It seems that 
the
conductivity tends to vanish as a power-law for small losses with an exponent close to
$1$ while it tends to saturate for larger losses with strong fluctuations in the region of the 
loss between $10^{-3}$ and $10^{-6}$ (while for smaller or
larger losses the conductivity seems to be self averaged). The behavior for small losses is 
in agreement with the theoretical predictions. Indeed,
for vanishing losses both dielectric and metallic components of the film are non-dissipative
and then the whole system becomes non-dissipative. Furthermore, from Eqs. (1-4), the 
real part
of the effective conductivity should behave as 
\begin{equation}
Real(\sigma_{eff}) = \frac{1}{2} \frac{R}{\omega} \end{equation}
which seems to be well fitted in Fig.3a. Although a difference is shown between the two 
methods, they follow qualitatively the same behavior for small losses.
However, as expected from Figs.2, the IPR shows a delocalization (for the RSRG method) 
for
increasing loss while the EM method yields the inverse situation which is the expected
behavior (for the same configuration, an increase of the loss corresponds to an increase of
the absorption which means a localization \cite{Gupt}). Therefore, this confirms the failure
of this renormalization group method in describing the localization properties of the system,
due to the large field strengths obtained by this method which broaden the distributions 
shown
in Fig.2. On the other hand, the IPR slope obtained by the EM method seems to saturate at 
the
value $-1.3$ in the region of small losses (see Fig.3b) indicating that the eigenmodes 
remain
localized even for vanishing losses. Indeed, this localization is due only to the disorder in 
the
conductivity (Anderson like localization). The disorder here does not come from the 
strength of the
conductivity (since the conductivities of the two components have the same strength) but 
from
its phase which takes randomly two values: $+\frac{\pi}{2}$ and $-\frac{\pi}{2}$ for 
vanishing losses. We  also see from this Figure, in
the region of loss between $10^{-3}$ and $10^{-6}$, a non monotonic behavior of the IPR 
(EM).
This is due to the strong fluctuations observed in this region. Therefore, this region 
should
show a different statistical behavior.

\newpage

\noindent{\bf Conclusion }

\hspace{0.33in}
In this article we have studied, by using the RSRG method and also the EM method, the 
localization
and absorption properties of the electromagnetic field in a thin semicontinuous 
metal-dielectric
film for a characteristic frequency $\omega_{res} =1$ at the classical percolation threshold. 
It
seems that the real part of the effective conductivity for  RSRG agrees qualitatively with 
the
exact calculations (EM) and give the expected power-law behavior for vanishing loss. 
While this
method (RSRG) fails in giving the right behavior for the IPR, due to the large field strengths
provided in comparison with the EM method. On the other hand, it seems that the IPR 
saturates at
$-1.3$ for vanishing losses in agreement with the anomalous experimental results observed 
for
such films. We can explain the anomalous absorption observed by a confinement of the 
light at
the percolation threshold. We also found  that the conductivity strongly fluctuates in the 
region
of the loss between $10^{-3}$ and $10^{-6}$ while it is self averaging for larger or 
smaller
losses. An extensive study of the statistical properties of the conductivity is then needed.
Indeed, may be in that region the logarithm of the conductivity is self averaged in this 
region.
Furthermore, it should be interesting to study these effective properties around 
the classical threshold where interesting features can occur. It is also important to examine
these effects for a partially ordered sample since realistic films show a local arrangement.
All these interesting features will be the subject of a forthcoming investigation.
\bigskip

\noindent{\bf Acknowledgments }

\hspace{0.33in} Two of us (L.Z and N.Z) would like to acknowledge the hospitality of the
ICTP during the progress of this work. This work was done within the framework of the Abdus
Salam International Centre for Theoretical Physics, Trieste, Italy. Financial support from the 
Arab Fund is acknowledged.

\newpage

\begin{center}
{\bf Figure Captions}
\end{center}

\bigskip

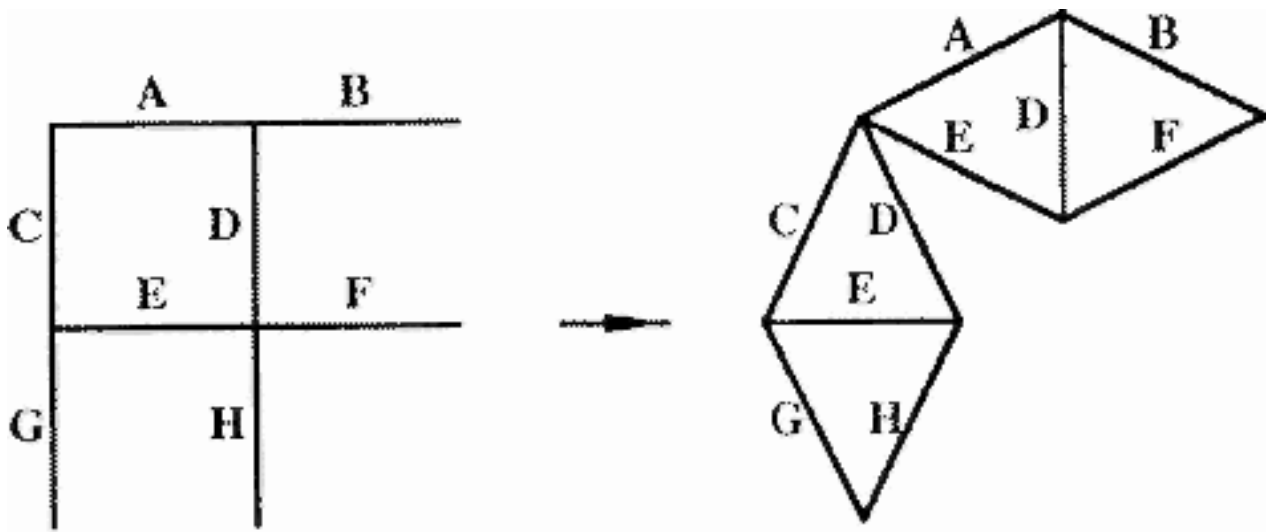
{\bf Fig.1 } The real space renormalization group for a square network. 

\bigskip

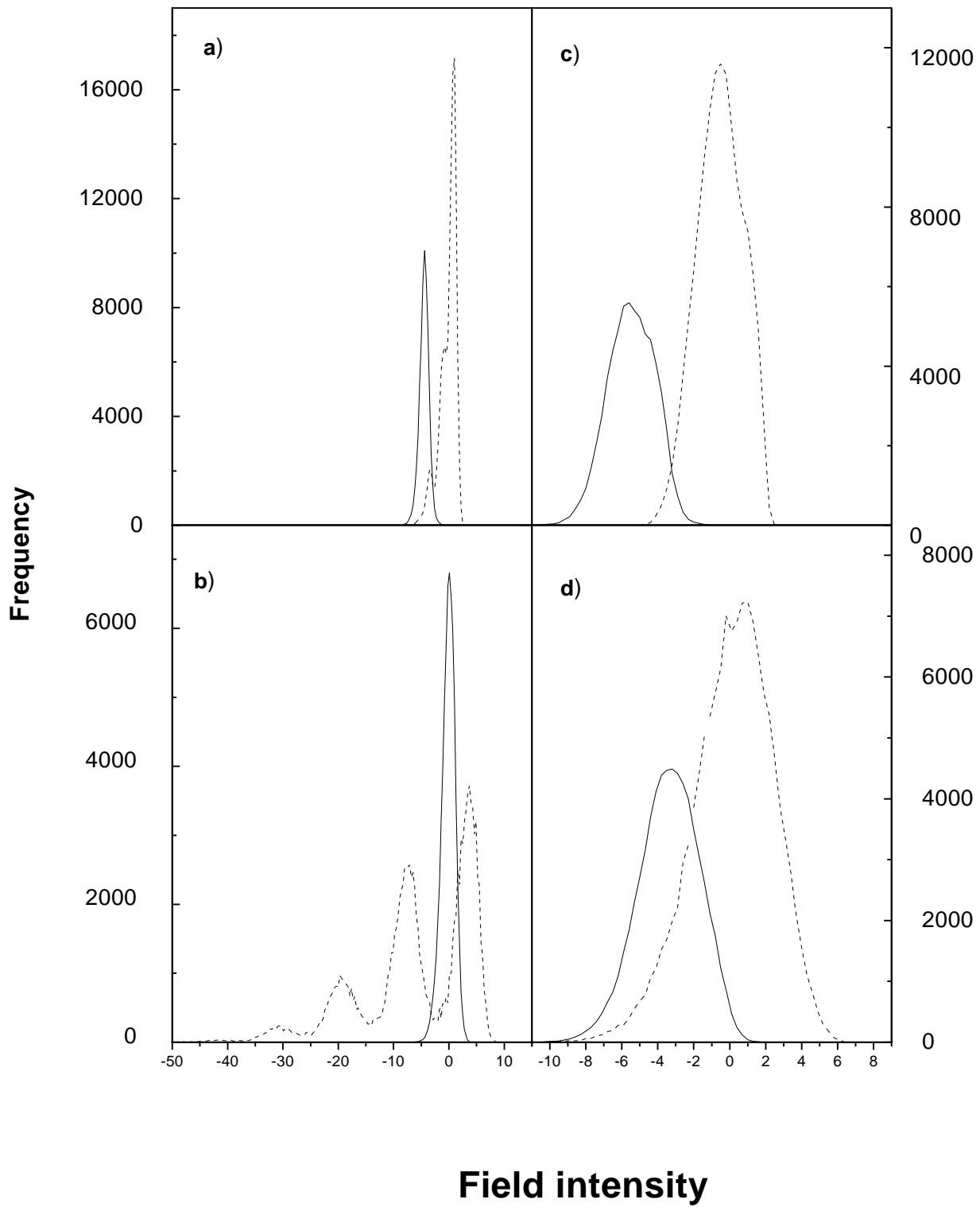
{\bf Fig.2} The distribution of the local electric field intensity $\rm log (\left| E \right| 
^{2})$ for $\omega =1$ : a) $ R=10^{-1}$, b) $R=10^{-6}$, and $\omega = 1/8$ : c) 
$R=10^{-1}$, d) $R=10^{-6}$. Solid curves correspond to the EM calculations and 
dashed curves to RSRG method.

\bigskip

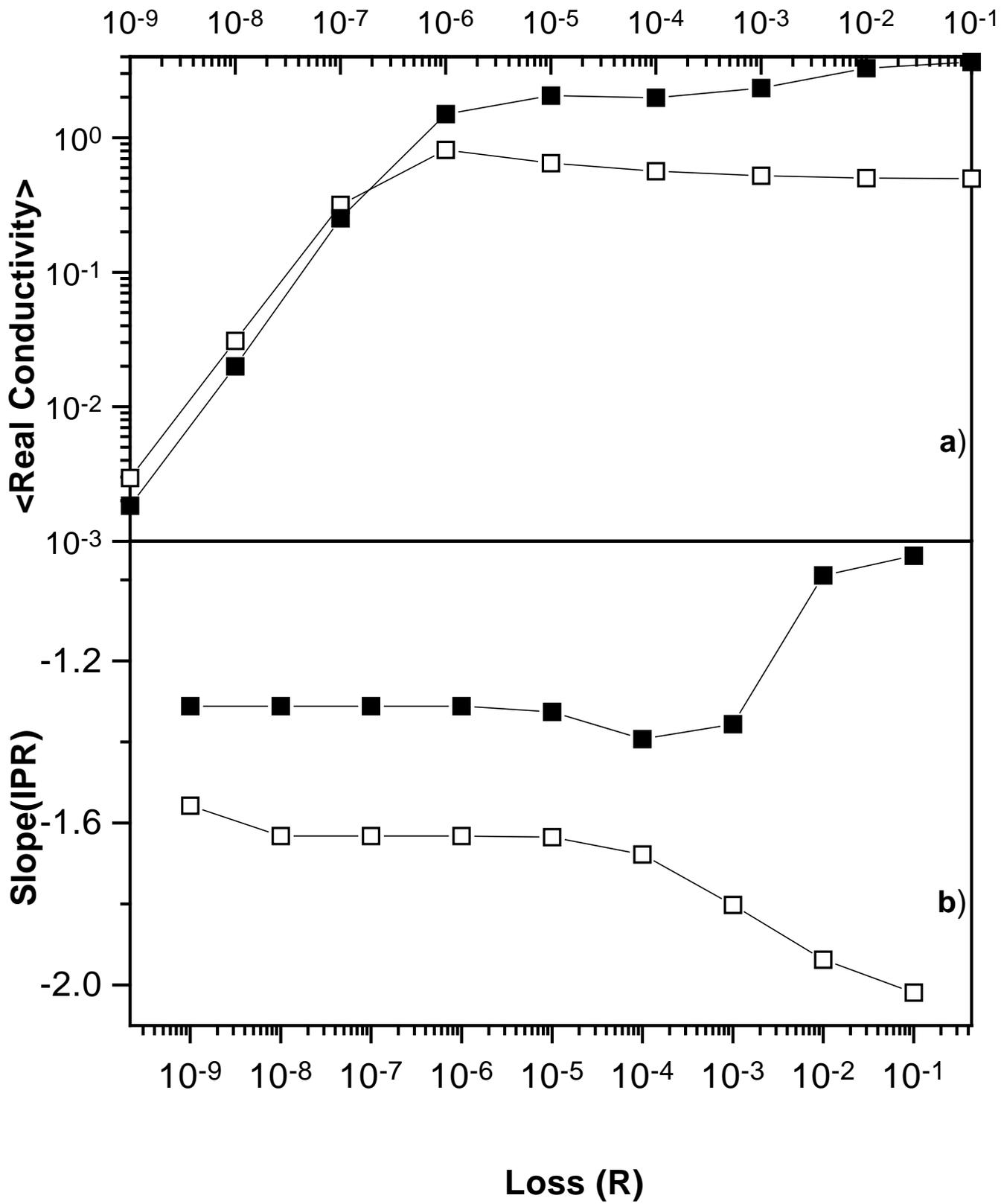
{\bf Fig.3} a) The real part of the effective conductivity in a log-log plot and b) the slope of 
the IPR in a semi-log plot as a function of the loss parameter $R$ for $\omega =1$. Open 
squares correspond to the RSRG method and solid squares to the EM method. The conductivity is 
averaged over 50 samples and the IPR is calculated for only one configuration.

\end{document}